\documentclass[aps,prb,showpacs,amsmath,amssymb,superscriptaddress,longbibliography,reprint,10pt]{revtex4-2}
\usepackage{graphicx}
\usepackage{xcolor}
\usepackage{newtxtext,newtxmath}
\usepackage{dsfont}
\usepackage{braket}

\usepackage[colorlinks=true,breaklinks=true,allcolors=blue]{hyperref}

\newcommand\bphi{\boldsymbol{\phi}}

\newcommand{\bfvec}[1]{\mathbf{#1}}
\newcommand{\ev}[1]{\langle #1 \rangle}
\DeclareMathOperator{\Tr}{{Tr}}
\DeclareMathOperator{\myIm}{{Im}}
\DeclareMathOperator{\myRe}{{Re}}

\newcommand{\cvec}[2]{\begin{bmatrix} #1 \\ #2 \end{bmatrix}} 
\newcommand{\rvec}[2]{\begin{bmatrix} #1 & #2 \end{bmatrix}} 
\newcommand{\mat}[2]{\begin{bmatrix} #1 \\ #2 \end{bmatrix}} 
\newcommand{\ul}[1]{{\underline{#1}}}

\makeatletter
\renewcommand*\env@matrix[1][\arraystretch]{%
  \edef\arraystretch{#1}%
  \hskip -\arraycolsep
  \let\@ifnextchar\new@ifnextchar
  \array{*\c@MaxMatrixCols c}}
\makeatother

\allowdisplaybreaks

\begin{document}

\title{Long-range Kitaev chain in a thermal bath:\texorpdfstring{\\}{} analytic techniques for time-dependent systems and environments}

\author{Emma C. King}
\affiliation{Institute of Theoretical Physics, University of Stellenbosch, Stellenbosch 7600, South Africa}

\author{Michael Kastner}
\affiliation{Institute of Theoretical Physics, University of Stellenbosch, Stellenbosch 7600, South Africa}
\affiliation{Hanse-Wissenschaftskolleg, Lehmkuhlenbusch 4, 27753 Delmenhorst, Deutschland}

\author{Johannes N. Kriel}
\affiliation{Institute of Theoretical Physics, University of Stellenbosch, Stellenbosch 7600, South Africa}

\date{\today}

\begin{abstract}
We construct and solve a ``minimal model'' with which nonequilibrium phenomena in many-body open quantum systems can be studied analytically under time-dependent parameter changes in the system and/or the bath. Coupling a suitable configuration of baths to a Kitaev chain, we self-consistently derive a Lindblad master equation which, at least in the absence of explicit time dependencies, leads to thermalization. Using the method of Third Quantization we derive time-evolution equations for the correlation matrix, which we relate to the occupation of the system's quasiparticle modes. These results permit analytic and efficient numeric descriptions of the nonequilibrium dynamics of open Kitaev chains under a wide range of driving protocols, which in turn facilitate the investigation of the interplay between bath-induced dissipation and the generation of coherent excitations by nonadiabatic driving. We advertise this minimal model of maximum simplicity for the study of finite-temperature generalizations of Kibble-Zurek ramps, Floquet physics, and many other nonequilibrium protocols of quantum many-body systems driven by time-varying parameters and/or temperatures.
\end{abstract}


\maketitle 

\section{Introduction}

Experimental advances continue to pave the way for probing the nonequilibrium dynamics of many-body quantum systems and for the realization of quantum-based technologies. While intriguing quantum effects are usually most strikingly on display in isolated systems, laboratory realizations of quantum systems will always be affected by some coupling to the environment. Hence, on the theory side, there is a need to complement the experimental developments with analytic and numeric techniques that provide insight into the nontrivial interplay between unitary and dissipative dynamics in open quantum systems at finite temperature.

The goal of this paper is to demonstrate how a combination of established techniques, augmented by model-specific adjustments, can give rise to faithful and exactly-solvable open-system versions of quadratic fermionic lattice models at finite temperature. To be specific, and for its wealth of physical phenomena to explore, we choose as a setting a long-range Kitaev chain \cite{Kitaev01,Vodola_etal16}. In isolation, this model of a one-dimensional topological superconductor has served as a setting for the investigation of topological phase transitions and topological order. Our aim is to derive for this model a Lindblad master equation that leads to thermalization at late times, so that temperature can be used as a control-knob of the system. Specifically, to explore a broad range of nonequilibrium phenomena, we want to permit time-dependent system Hamiltonians as well as time-dependent bath temperatures. This will allow us to study not only finite-temperature effects in Floquet or Kibble-Zurek protocols and the concomitant effects on topological properties of the Kitaev chain, but additionally opens up possibilities to study nonequilibrium quantum dynamics under temperature quenches, temperature ramps, or temperature oscillations.

We start from a microscopic picture in which each site of the Kitaev chain is weakly coupled to its own independent thermal bath, with all the baths characterized by the same temperature. We employ a procedure by D'Abbruzzo and Rossini \cite{DAbbruzzoRossini21}, tailored to quadratic Hamiltonians, which allows us to derive the corresponding Lindblad master equation in a self-consistent manner. In the absence of explicit time-dependencies in the Hamiltonian and the baths, this construction ensures thermalization of the Lindblad equation at late times. This is in contrast to previous studies of nonequilibrium dynamics of open Kitaev chains \cite{Keck_etal17,NigroRossiniVicari19,RossiniVicari20}, where Lindblad equations with local jump operators were postulated {\em ad hoc}, which may be reasonable choices for the modeling of dissipative effects at shorter times, but fail to thermalize at late times.

The Lindblad master equation we derive, albeit nonlocal in the jump operators, is quadratic in the fermionic operators. Lindbladians of this type can be solved analytically by a diagonalization technique termed Third Quantization \cite{Prosen08}. Going beyond this general technique, we can exploit specific properties of the Kitaev chain and the thermal environment to further simplify the solution, which further increase the accessible system sizes when evaluating the resulting expressions on a computer. In addition to providing eigenstates and eigenvalues of the Lindbladian, the formalism of Third Quantization also provides a convenient setting for studying the dynamics of the correlations between fermionic operators \cite{KosProsen17}. Here, exploiting again specific properties of the Lindbladian, most notably its translational invariance, the time evolution of the correlation matrix can be broken down into $4\times4$ blocks. Via Wick's theorem, the resulting matrix differential equations give access to the dynamics of arbitrary correlation functions and mode occupation numbers of the open Kitaev chain.

As a first illustration of the usefulness of our results, we calculate the time evolution of the mode occupation numbers of a Kitaev chain with nearest-neighbor hopping and pairing which is in contact with a thermal environment at positive temperature and for which a parameter in the Hamiltonian is ramped (increased linearly in time). Via numerical integration of the time-evolution equations of the mode occupation numbers we obtain data for system sizes of several thousand lattice sites. Plots of these data show the pronounced creation of excitations when ramping the parameter across the critical parameter value of the quantum phase transition, which can be interpreted as a finite-temperature manifestation of the quantum Kibble-Zurek effect \cite{Patane_etal08}. The simplicity of our model not only allows for the efficient numeric treatments of large systems, but also, in certain limits, for the derivation of a rate equation for mode occupation numbers, similar to the one postulated for a spin model in Ref.\ \cite{Patane_etal08}. Based on this rate equation, scaling laws governing the excitation density at the end of the ramp can be derived analytically. The full extent of the power and versatility of our results is on display in the companion paper ``Universal cooling dynamics towards a quantum critical point,'' in which, based on the analytic results for the time-evolution of mode occupation numbers, Kibble-Zurek-type scaling laws for protocols with time-dependent {\em temperatures} are derived \cite{KingKastnerKrielLetter}.

The paper is organized as follows. In Sec.~\ref{s:Kitaev} we introduce the long-range Kitaev chain, outline the procedure through which it can be diagonalized, and provide an overview of the model's zero-temperature phase structure. Section~\ref{s:OpenKitaev} deals with the coupling of the chain to a thermal environment, starting from an interacting system--bath Hamiltonian to eventually obtain a Lindblad master equation. This equation is treated using the method of Third Quantization in Sec.~\ref{s:diagonaliseLindbladian}. Section~\ref{s:secTimeEvolutionCorrelations} is dedicated to deriving the equations of motion for fermionic correlations, and relating these to the occupation of the system's quasiparticle modes. In Sec.~\ref{s:example} we present results for a particular parameter ramp across the system's critical point, performed at a fixed finite temperature. Here we highlight the competing processes which emerge both in the numeric results and in suitably adapted versions of the dynamic equations. A summary and conclusions follow in Sec.~\ref{s:summary}. 


\section{Long-range Kitaev chain}
\label{s:Kitaev}
We introduce a version of the Kitaev chain containing long-range hopping and pairing terms, and outline the procedure through which this model's Hamiltonian can be diagonalized. Aided by these results, we also give a brief overview of the model's zero temperature phase structure. While the results of this section are fairly standard and can be found in the literature, the diagonalized Hamiltonian and the corresponding linear transformations will be useful when dealing with the open-system version of the Kitaev chain in Sec.~\ref{secThirdQuantizationKitaev}.

\subsection{Model definition}
\label{ss:Kitaev Definition}
The long-range Kitaev chain is defined by the Hamiltonian~\cite{Vodola_etal16,DuttaDutta17}
\begin{align}\label{eq:LRKitaevChainHamiltonian}
				H_\text{S} =&  \sum_{j=1}^L{\sum_{l=1}^{\left\lfloor L/2 \right\rfloor}{\left[\left(J d_l^{-\phi} c_j^\dagger c_{j+l}^{\phantom\dag} + \frac{\Delta}{2}d_l^{-\alpha}c_j c_{j+l}\right) + \text{h.c.} \right]}} \nonumber
				\\&+ 2 \mu \sum_{j=1}^L{c_j^\dagger c_j^{\phantom\dag}},
\end{align}
where the operators $c^\dag_j$ and $c_j$ create, respectively destroy, a spinless fermion at site $j$ on a chain with $L$ sites. We enforce periodic boundary conditions, which generate a translationally invariant closed-ring geometry. The parameters $J$ and $\Delta$ appearing in $H_\text{S}$ are the hopping and pairing strengths, and $\mu$ is the on-site chemical potential. The parameters $\phi$ and $\alpha$ govern the power-law decay of the hopping and pairing terms~\cite{DuttaDutta17} with the distance
\begin{align}
d_l = \begin{cases}
l & \text{if}\ \ l<L/2,\\
2^{1/\phi}l & \text{if}\ \ l = L/2.
\end{cases}
\label{eq:effectiveDistanceBetweenLatticeSites}
\end{align}
The factor of $2^{1/\phi}$ in $d_{L/2}$ only survives in the hopping terms, and prevents an overcounting of terms with a hopping range of $l=L/2$. In the limit $\phi,\alpha \rightarrow \infty$ we recover Kitaev's original model with nearest-neighbor hopping and pairing \cite{Kitaev01}. We restrict the discussion to systems where $\phi,\alpha>1$ in order to avoid complications related to nonextensivity and large-system limits.

\subsection{Diagonalization}
\label{ss:Kitaev Diagonalization}
The Kitaev chain is an exactly solvable many-body system and can be diagonalized through a combination of Fourier and Bogoliubov transformations; see Refs.~\cite{Lieb61,Vodola_etal16,DuttaDutta17,DAbbruzzoRossini21} for more comprehensive accounts. We introduce Fourier transforms
\begin{equation}
a_k = \frac{1}{\sqrt{L}}\sum_{j}{e^{ijk}c_{j}}
\end{equation}
of the fermionic site operators, which allow the Hamiltonian~\eqref{eq:LRKitaevChainHamiltonian} to be recast as
\begin{equation}
			H_\text{S} = \frac{1}{2} \sum_{k}\psi_k^\dag \mathbf{H}_\text{BdG}(k) \psi_k
			\label{eq:lrkMatrixForm}
\end{equation}
with $\psi_k^\dag = \rvec{a_{k}^\dagger}{a_{-k}}$. Here the lattice momenta $k\in\{2\pi n/L:n\in\mathbb{Z}_L\}$ run over the first Brillouin zone. The $2 \times 2$ Bogoliubov--de Gennes (BdG) Hamiltonian in Eq.~\eqref{eq:lrkMatrixForm} reads
\begin{equation}
			\mathbf{H}_\text{BdG}(k) = \mat{2(J g_\phi(k) + \mu) & i\Delta f_\alpha(k)}{
				-i\Delta f_\alpha(k) & - 2(J g_\phi(k) + \mu)},
			\label{eq:BdGHamiltonian}
\end{equation}
where we have introduced
\begin{equation}
g_\phi(k) = \sum_{l=1}^{\left\lfloor L/2 \right\rfloor}{d_l^{-\phi}\cos(k l)},\quad f_\alpha(k) = \sum_{l=1}^{\left\lfloor L/2 \right\rfloor}{d_l^{-\alpha}\sin(k l)}.
\end{equation} 

Next we utilize the Bogoliubov transformation
			\begin{equation}
			\cvec{\eta_{k}^\dagger}{
				\eta_{-k}}
			= \mat{
				\cos \beta_{k} & i \sin \beta_{k}}{
				i \sin \beta_{k} & \cos \beta_{k}}
			\cvec{
				a_{k}^\dagger}{
				a_{-k}}
			\label{eq:bogo}
			\end{equation}
with the Bogoliubov angle $\beta_k$ set by
\begin{equation}
\tan(2\beta_{k})=-\Delta f_\alpha(k)/(2J g_\phi(k)+2\mu).
\end{equation}
In terms of the $\eta_k$ fermions defined in Eq.\ \eqref{eq:bogo} the Hamiltonian takes the diagonal form 
			\begin{equation}
					H_\text{S} = \sum_{k}{\lambda_{\phi,\alpha} (k) \left(\eta_{k}^\dagger \eta_{k}^{\phantom\dag} -\tfrac{1}{2}\right)}
			\label{eq:kitaevChainHamiltonianDiagonalForm}
			\end{equation}
with dispersion relation
\begin{equation}
\lambda_{\phi,\alpha}(k) = \sqrt{[2J g_\phi(k)+2\mu]^2 + [\Delta f_\alpha(k)]^2}.
\label{eq:dispersionRelationOriginal}
\end{equation} 

In the derivation of the Lindblad master equation in Sec.~\ref{s:OpenKitaev} it will be convenient to combine the Fourier and Bogoliubov transformations into a single canonical transformation which relates the $c_j$ site fermions directly the $\eta_k$ fermions of the quasiparticle modes. By organizing the site operators into column vectors $\ul{c} = [c_1,\dotsc,c_{L}]^\text{T}$ and doing the same for the operators of the quasiparticle modes, the combined transformation can be written as \cite{DAbbruzzoRossini21}
\begin{equation} \label{eq:canonicalTransformation}
\cvec{\ul{c}}{\ul{c}^{\,\dagger}}
	 = \mathbf{T} \cvec{\ul{\eta}}{\ul{\eta}^{\,\dagger}}\quad\text{where}\quad\mathbf{T}=\mat{\mathbf{V}&\mathbf{S}}{\mathbf{S}^*&\mathbf{V}^*}.
\end{equation}
Here, $\mathbf{V}$ and $\mathbf{S}$ are $L\times L$ matrices with entries 
\begin{equation}\label{e:VS}
	\mathbf{V}_{jk}=\frac{1}{\sqrt{L}} e^{-ijk} \cos \beta_k,\qquad \mathbf{S}_{jk}=\frac{-i}{\sqrt{L}}e^{+ijk}\sin\beta_k.
\end{equation}
In general, for the $\mathbf{T}$ transformation to preserve the fermionic anticommutation relations it is required that
\begin{subequations}
\begin{align}
\mathbf{V}\mathbf{V}^\dagger + \mathbf{S}\mathbf{S}^\dagger &= \mathbf{V}^\dagger \mathbf{V} + \mathbf{S}^\text{T}\mathbf{S}^*= \boldsymbol{\mathbb{I}},\label{eq:conditionsOnVandSmatricesAntiCOmRelationsA}\\
\mathbf{V}\mathbf{S}^\text{T} + \mathbf{S}\mathbf{V}^\text{T} &= \mathbf{S}^\text{T}\mathbf{V}^* + \mathbf{V}^\dagger \mathbf{S} = \mathbf{0}.
\label{eq:conditionsOnVandSmatricesAntiCOmRelationsB}
\end{align}
\end{subequations}
It is straightforward to verify that $\mathbf{V}$ and $\mathbf{S}$ indeed satisfy these conditions.

\subsection{Phase structure}
\label{ss:Kitaev Quantum Phases}
\begin{figure}%
\includegraphics[width=0.8\columnwidth]{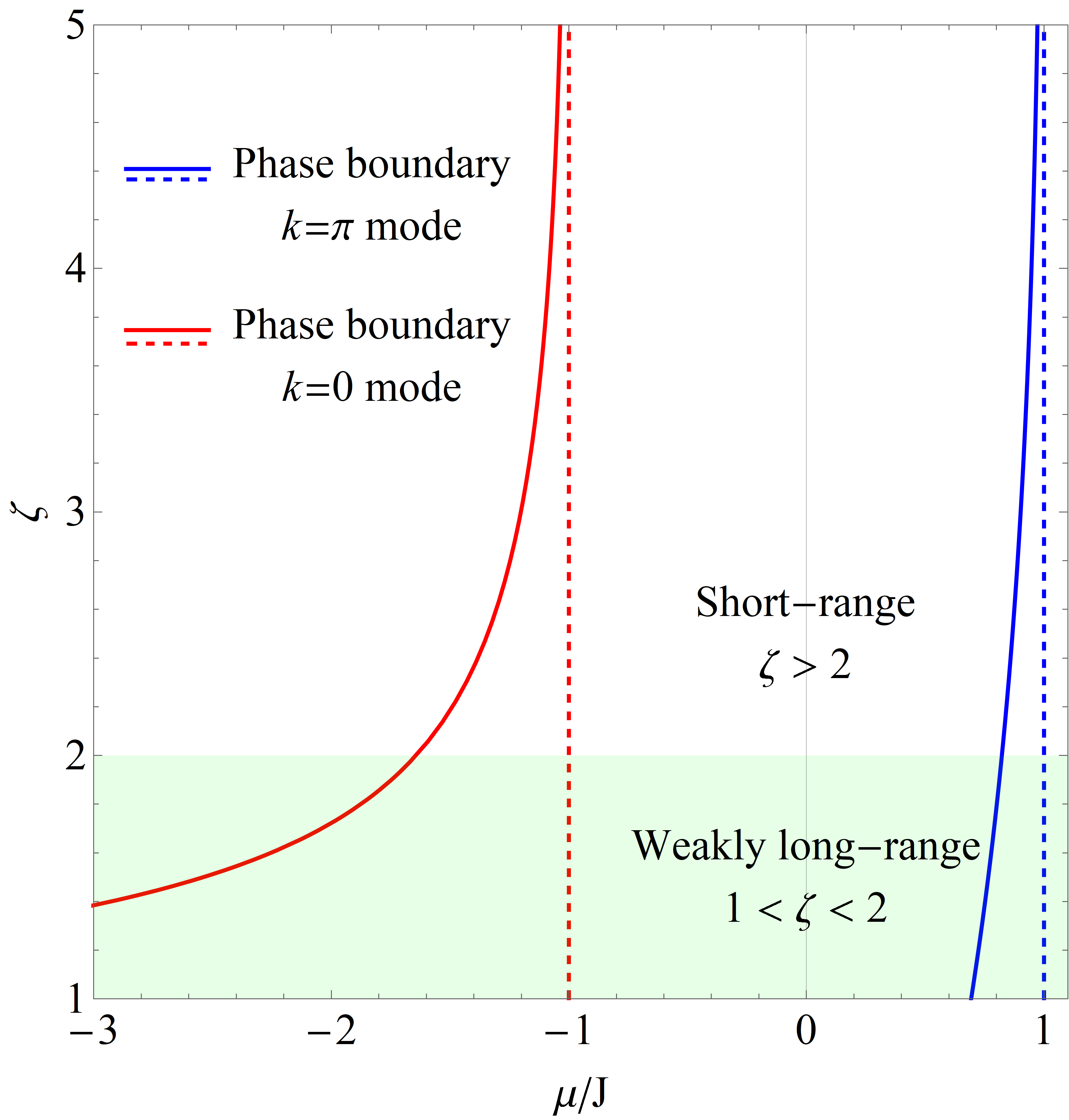}%
\caption{The phase boundaries of the Kitaev chain in the $(\zeta,\mu/J)$ plane for a system size of $L=1500$. Here $\zeta$ denotes one of the two long-range parameters, either $\alpha$ or $\phi$. Solid lines separate the phases of the Kitaev chain with nearest-neighbor pairing and arbitrary range hopping for which $\alpha=\infty$ and $\zeta=\phi>1$.  Dashed lines correspond to the phase boundaries for a chain with nearest-neighbor hopping $\phi=\infty$ and arbitrary range pairing $\zeta=\alpha>1$.}%
\label{fig:phaseBoundary}%
\end{figure}

The Kitaev chain exhibits two second-order quantum phase transitions at zero temperature. These transitions are signaled by the closure of the system's excitation gap, which occurs when the mode energy $\lambda_{\phi,\alpha}(k)$ in Eq.~\eqref{eq:dispersionRelationOriginal} vanishes for a particular momentum $k$. For this to happen,
\begin{equation}
f_\alpha(k)=0, \qquad g_\phi(k) = -\mu/J
\label{eq:conditionsCloseGap}
\end{equation}
have to hold simultaneously. To determine the phase boundaries, we note that $f_{\alpha>1}(k)$ vanishes only at $k=0,\pi$. Disregarding the null-pairing case, this reveals the existence of two critical lines in the $(\phi,\mu/J)$ or $(\alpha,\mu/J)$ planes: one corresponding to $g_\phi(0) = -\mu/J$ and the other to $g_\phi(\pi) = -\mu/J$, as shown in Fig.~\ref{fig:phaseBoundary}. In this figure as well as in the discussion that follows, either the hopping or pairing are taken to be nearest-neighbor processes, i.e., either $\phi=\infty$ or $\alpha=\infty$. We have split the phase diagram in Fig.~\ref{fig:phaseBoundary} into two regions, short-range and weakly long-range, with short-range referring to hopping and pairing exponents $\phi>2$ and $\alpha>2$, respectively. Weakly long-range systems are those for which $\phi\in(1,2)$ or $\alpha\in(1,2)$. For nearest-neighbor hopping $(\phi=\infty)$ and arbitrary range pairing $(\alpha>1)$ we have $g_\infty(k)=\cos(k)$, and so the phase boundaries are always located at $\mu/J=\mp 1$. For nearest-neighbor pairing $(\alpha=\infty)$ and arbitrary range hopping $(\phi>1)$ the more complicated behavior of $g_\phi(k)$ results in the phase boundaries deviating from $\mu/J=\mp 1$ at finite $\phi$.

\section{Kitaev chain coupled to a bath}
\label{s:OpenKitaev}
\begin{figure}
\centering
\includegraphics[width=\columnwidth]{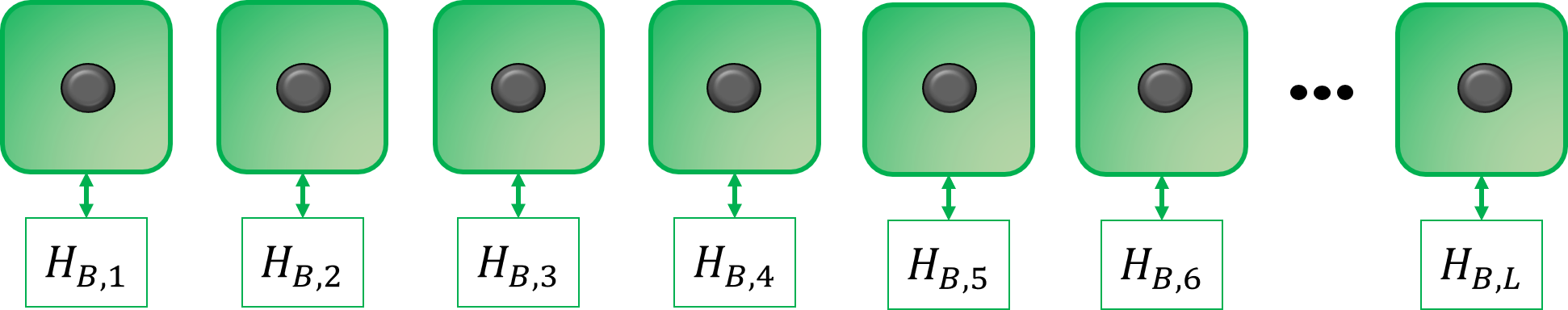}%
\caption{Schematic diagram of the system--environment interaction, where the circles represent sites on the chain and the squares are the independent bosonic baths. Each lattice site is coupled to its own bath, and all the baths together compose the environment.}%
\label{fig:schematicSystemBathCoupling}%
\end{figure}

Starting from system and bath Hamiltonians, we derive a Lindblad master equation describing the open-system dynamics of the Kitaev chain. We start from a microscopic system--bath configuration where each site on the chain is coupled to an individual bosonic bath. These baths all have the same temperatures and are also otherwise identical. This is depicted schematically in Fig.~\ref{fig:schematicSystemBathCoupling}. Together, the chain and baths form an isolated quantum system which evolves unitarily under the Hamiltonian
\begin{equation}
H = H_\text{S} \otimes \mathbb{I}_\text{E} + \mathbb{I}_\text{S} \otimes H_\text{E} + H_{\text{int}},
\label{eq:totalIsolatedSystem}
\end{equation}
where $H_\text{S}$ is the system Hamiltonian~\eqref{eq:kitaevChainHamiltonianDiagonalForm} and $\mathbb{I}_\text{S}$ and $\mathbb{I}_\text{E}$ are identity operators on the Hilbert spaces of the system and the environment, respectively. The environment, consisting of $L$ identical thermal baths with mode frequencies $\omega_{q}$, is defined by
\begin{equation}
	H_\text{E} = \sum_{n=1}^{L}{H_{\text{B},n}}, \qquad H_{\text{B},n} = \int dq \, \omega_{q} b_{n,q}^\dagger b_{n,q}.
\end{equation}
Here $b_{n,q}$ is the bosonic operator associated with mode $q$ of the $n$th bath. 
For the system--environment interaction we choose a linear coupling between the bath modes and sites on the chain,
\begin{equation}
H_{\text{int}} = \sum_{n=1}^{L} O_n \otimes R_n
 \label{eq:interactionHamiltonianBathCoupling}
\end{equation}
with
\begin{equation}
	O_n=(c_n + c_n^\dagger),\qquad R_n=\int dq \, g_{q} (b_{n,q}+ b_{n,q}^\dagger).
	\label{eq:OnRnDefinitions}
\end{equation}
 The coupling strength between bath mode $q$ and the $n$th lattice site is denoted $g_{q}$.

\subsection{Markovian master equation}
The density operator of the system--baths composite evolves unitarily under the Hamiltonian \eqref{eq:totalIsolatedSystem} according to the von Neumann equation. Assuming that the baths are prepared in thermal equilibrium states at a temperature $T$, and that their coupling to the system has a negligible effect on their dynamics, it is possible to trace out the environment's degrees of freedom. Following a standard series of approximations, the chain's dynamics is described by the master equation \cite{BreuerPetruccione}
			\begin{equation}
					\frac{\text{d}\rho(t)}{\text{d}t}=  - i [H_\text{S} + H_\text{LS},\rho(t)] + \hat{\mathcal{D}}[\rho(t)],
					\label{eq:lindbladMasterEquationDiagonalChap2}
			\end{equation}
where $\rho$ is the density operator of the chain and $H_\text{LS}$ is the Lamb-shift Hamiltonian. The dissipation superoperator $\hat{\mathcal{D}}$ on the right-hand side of the master equation reads \cite{DAbbruzzoRossini21,BreuerPetruccione}
\begin{multline}
\hat{\mathcal{D}}[\rho] = \sum_{m,n,\lambda} \bar{\Gamma}_{mn}(\lambda)\\
\times\left[2O_n(\lambda) \rho O_m^\dagger(\lambda) - \left\{ O_m^\dagger(\lambda) O_n(\lambda),\rho\right\} \right],
\label{eq:dissipatorGeneralForm}
\end{multline}
where $m$ and $n$ are site labels and $\lambda$ ranges over all energy differences between $H_{\rm S}$ eigenvalues. Here,
\begin{equation}\label{eq:fourierTransformGeneral}
\bar{\Gamma}_{mn} (\lambda) = \frac{1}{2}\int_{-\infty}^\infty d\tau \, e^{i\lambda \tau} \langle \tilde{R}^\dagger_m(\tau)R_n \rangle
\end{equation}
with
\begin{equation}
  \tilde{R}_m(\tau)=e^{iH_\text{E} \tau}R_m e^{-iH_\text{E} \tau}
\end{equation}
is the Fourier transform of an environment correlation function. The remaining elements on the right-hand side of Eq.\ \eqref{eq:dissipatorGeneralForm} are the operators obtained from decomposing the operators $O_n=(c_n + c_n^\dagger)$ in Eq.~\eqref{eq:OnRnDefinitions} into eigenoperators of the system Hamiltonian,
\begin{equation}
O_n(\lambda)= \sum_{\epsilon'-\epsilon=\lambda} \Pi (\epsilon) O_n \Pi(\epsilon')
\label{eq:eigenoperatorsGeneral}
\end{equation}
where $\epsilon$ denotes an eigenvalue of $H_\text{S}$ and $\Pi(\epsilon)$ is the projection operator onto the corresponding eigenspace \cite{BreuerPetruccione}.

\subsection{Calculation of the dissipator and jump operators} \label{sec:dissipator}
Next we follow the procedure described in sections IV.B and IV.C of Ref.\ \cite{DAbbruzzoRossini21} to simplify the dissipator \eqref{eq:dissipatorGeneralForm} and to bring the master equation \eqref{eq:lindbladMasterEquationDiagonalChap2} into its final form. Starting with the environment correlation functions \eqref{eq:fourierTransformGeneral} and following step by step the procedure of Sec.~IV.C of Ref.~\cite{DAbbruzzoRossini21}, we find that $\bar{\Gamma}_{mn} (\lambda)=\delta_{mn}\bar{\Gamma}(\lambda)$ with 
\begin{align}
\bar{\Gamma}(\lambda) = \begin{cases}
 \mathcal{J}(-\lambda) n_\text{\tiny BE}(-\lambda) &\text{if $\lambda<0$},\\
\mathcal{J}(0) (2n_\text{\tiny BE}(0)+1) &\text{if $\lambda=0$},\\
\mathcal{J}(\lambda) (n_\text{\tiny BE}(\lambda)+1) &\text{if $\lambda>0$},
\end{cases}
\label{eq:gammaBarItoLambda}
\end{align}
where $n_{\text{\tiny BE}}(\lambda)=\left(e^{\lambda/T} - 1\right)^{-1}$ is the Bose-Einstein distribution and $\mathcal{J}(\lambda) = \pi \int dk \, |g_{k}|^2 \delta(\lambda-\omega_{k})$ is the density of states of a single bath. This result for $\bar{\Gamma}_{mn} (\lambda)$ brings the dissipator \eqref{eq:dissipatorGeneralForm} into the form
\begin{equation}
\hat{\mathcal{D}}[\rho] = \sum_{n,\lambda} \bar{\Gamma}(\lambda)\left[2O_n^{\phantom{\dag}}(\lambda) \rho O_n^\dagger(\lambda) - \left\{ O_n^\dagger(\lambda) O_n^{\phantom{\dag}}(\lambda),\rho\right\} \right].
\label{eq:dissipatorEq1}
\end{equation}
The eigenoperators $O_n(\lambda)$ in Eq.~\eqref{eq:eigenoperatorsGeneral} can be found by relating the site operators to those of the quasiparticle modes via the linear transformation in Eq.~\eqref{eq:canonicalTransformation}. This leads to \cite{DAbbruzzoRossini21}
\begin{equation}
O_n(\lambda) = \sum_k \left( \bphi_{nk}^{\phantom{\dag}}\delta_{\lambda,\lambda_k}^{\phantom{\dag}}\eta_k^{\phantom{\dag}} + \bphi_{nk}^*\delta_{\lambda,-\lambda_k}^{\phantom{\dag}}\eta_k^\dagger \right),
\label{eq:systemEigenoperator}
\end{equation}
where $\boldsymbol{\phi} = \mathbf{V} + \mathbf{S}^*$, with $\mathbf{V}$ and $\mathbf{S}$ as defined in Eq.~\eqref{e:VS}. To use Eq.\ \eqref{eq:systemEigenoperator} in order to further simplify the dissipator \eqref{eq:dissipatorEq1} and eventually extract the jump operators, we need to keep track of the degeneracies between the energies of the $\eta_k$ and $\eta_{-k}$ quasiparticle modes. Assuming for the moment that there are no zero-energy modes, we can insert Eq.~\eqref{eq:systemEigenoperator} into \eqref{eq:dissipatorEq1} to produce
\begin{align}
\hat{\mathcal{D}}[\rho] = \sum_{n,k} &\sum_{q=\pm k} \bigg[\Phi_{kq}^{n}\bar{\Gamma}(\lambda_k)\left(2 \eta_k\rho \eta_q^\dagger-\left\{\eta_q^\dagger \eta_k,\rho\right\}\right) \nonumber \\
&+ \Phi_{qk}^{n}\bar{\Gamma}(-\lambda_k)\left(2 \eta_k^\dagger\rho \eta_q- \left\{\eta_q \eta_k^\dagger,\rho\right\}\right) \bigg]\label{eq:dissipatorForDegeneracies}
\end{align}
with $\Phi_{kq}^{n} = \Phi_{qk}^{n*} = \bphi_{nk}^{\phantom{*}}\bphi_{nq}^*$. Further simplifications arise through constraints on the coefficients $\Phi_{kq}^{n}$, which are inherited from constraints on the matrices $\mathbf{S}$ and $\mathbf{V}$ in Eq.~\eqref{e:VS} and enter via $\bphi = \mathbf{V} + \mathbf{S}^*$. In addition to the conditions \eqref{eq:conditionsOnVandSmatricesAntiCOmRelationsA} and \eqref{eq:conditionsOnVandSmatricesAntiCOmRelationsB} from the Bogoliubov transformation, the particle-hole and time-reversal symmetries of the Kitaev Hamiltonian \eqref{eq:LRKitaevChainHamiltonian} impose the condition
\begin{equation}
\mathbf{V}^\dagger \mathbf{S}^* + \mathbf{S}^\text{T}\mathbf{V} = \mathbf{0}.
\end{equation}
See Appendix \ref{app:symmetries} for details. Overall these conditions ensure that $\bphi$ is unitary,
\begin{equation}
\bphi^\dagger \bphi = (\mathbf{V}^\dagger + \mathbf{S}^\text{T})(\mathbf{V}+\mathbf{S}^*) = \mathbb{I}.
\label{eq:phiUnitaryCondition}
\end{equation}
It now follows that
\begin{equation}
\sum_n \Phi_{kq}^{n} = (\bphi^\dagger \bphi)^*_{kq} = \delta_{kq},
\end{equation}
which reduces the dissipator in Eq.~\eqref{eq:dissipatorForDegeneracies} to
\begin{align}
\hat{\mathcal{D}}[\rho]= \sum_{k} &\bigg[\bar{\Gamma}(\lambda_k)\left(2 \eta_k^{\phantom{\dagger}}\rho \eta_k^\dagger-\left\{\eta_k^\dagger \eta_k^{\phantom{\dagger}},\rho\right\}\right)\nonumber \\
&+ \bar{\Gamma}(-\lambda_k)\left(2 \eta_k^\dagger\rho \eta_k^{\phantom{\dagger}}- \left\{\eta_k^{\phantom{\dagger}} \eta_k^\dagger,\rho\right\}\right) \bigg].\label{eq:dissipatorSimplifiedForm}
\end{align}
To finally extract the jump operators, we combine the master equation \eqref{eq:lindbladMasterEquationDiagonalChap2} and the simplified dissipator \eqref{eq:dissipatorSimplifiedForm} to obtain the master equation for the Kitaev chain in Lindblad form as
			\begin{align}
					\dot{\rho}=& - i \left[H_\text{S} + H_\text{LS},\rho\right]\nonumber \\
					&+ \gamma \sum_k \sum_{\sigma=\pm} \left(2L_{k,\sigma}^{\phantom{\dagger}}\rho L_{k,\sigma}^\dagger -\left\{L_{k,\sigma}^\dagger L_{k,\sigma}^{\phantom{\dagger}},\rho\right\}\right) \label{eq:lindbladMasterEquationDiagonal}.
			\end{align}
Here we have introduced $\gamma$, previously contained in the $g_q$ coefficients in Eq.~\eqref{eq:OnRnDefinitions}, as a parameter for controlling the system--bath coupling strength. The jump operators in Eq.~\eqref{eq:lindbladMasterEquationDiagonal} are read off as
		\begin{equation}
				L_{k,+} = \sqrt{\Gamma_{k,+}}\eta_{k}^\dagger, \qquad L_{k,-} = \sqrt{\Gamma_{k,-}}\eta_{k},
						\label{eq:LindbladBathOperators}
		\end{equation}
with bath coupling constants
\begin{subequations}
\begin{align}
 \Gamma_{k,+}&= \bar{\Gamma}(-\lambda_k) = \mathcal{J}(\lambda_k) n_\text{\tiny BE}(\lambda_k),\label{eq:LindbladBathCouplingConstants1}\\
 \Gamma_{k,-} &= \bar{\Gamma}(\lambda_k) = \mathcal{J}(\lambda_k) \left(n_\text{\tiny BE}(\lambda_k)+1\right).
\label{eq:LindbladBathCouplingConstants2}
\end{align}
\end{subequations}
The first term of Eq.~\eqref{eq:lindbladMasterEquationDiagonal} contains the  Lamb-shift correction $H_\text{LS}$, which amounts to a bath-induced modification of the system's unitary dynamics \cite{BreuerPetruccione,DAbbruzzoRossini21}. In Appendix~\ref{app:LSHamiltonian} we show that $H_\text{LS}$ is diagonal in the $\eta_k$ fermions and hence does not introduce a coupling between these modes, but only a shift proportional to $\gamma$ in the mode energies in the unitary part of the master equation. Since the system--bath coupling strength $\gamma$ is weak, we will neglect the contribution of $H_\text{LS}$ in what follows. In the derivation of Eqs.~\eqref{eq:lindbladMasterEquationDiagonal}--\eqref{eq:LindbladBathCouplingConstants2} we assumed that the energies of the quasiparticle modes are all strictly positive. As discussed in Sec.~\ref{ss:Kitaev Quantum Phases}, this is not the case at one of the critical points of the Kitaev chain, where the energy of either the $k=0$ or $k=\pi$ mode vanishes. If, for example, $\lambda_0=0$ then, through a similar series of steps as before, we would obtain a modified $k=0$ contribution to the dissipator in Eq.~\eqref{eq:dissipatorSimplifiedForm} which is of the form
\begin{equation}
\hat{\mathcal{D}}_0[\rho] = \bar{\Gamma}(0)\left[2\left(\eta_0^{\phantom{\dagger}}+\eta_0^\dagger\right)\rho\left(\eta_0^{\phantom{\dagger}}+\eta_0^\dagger\right)-\left\{\left(\eta_0^{\phantom{\dagger}}+\eta_0^\dagger\right)^2,\rho\right\}  \right].
\label{eq:zeroModeDissipator}
\end{equation}
This suggests that there is now a single jump operator $L_0 = \sqrt{2\bar{\Gamma}(0)}(\eta_0^{\phantom{\dagger}}+\eta_0^\dagger)$ associated with this mode. Despite this, it can be shown that the different structures of the $k=0$ terms in Eqs.~\eqref{eq:dissipatorSimplifiedForm} and \eqref{eq:zeroModeDissipator} do not impact on the dynamics of the mode occupation $\langle \eta^\dagger_0 \eta_0^{\phantom{\dagger}} \rangle$, which is our primary quantity of interest in later applications. For our purposes, the Lindblad master equation \eqref{eq:lindbladMasterEquationDiagonal} will therefore suffice to treat all the quasiparticle modes, including the zero-energy mode at a critical point.

\section{Diagonalizing the Lindbladian in Liouville space} \label{s:diagonaliseLindbladian}
Prosen developed in Ref.~\cite{Prosen08} a general technique for solving Lindblad master equations in which the Hamiltonian is quadratic and the jump operators are linear in the fermionic operators. This approach makes use of a canonical quantization procedure in Liouville space and goes under the name of Third Quantization. We establish some groundwork for this method in Sec.~\ref{secThirdQuantization} and then apply it to the master equation \eqref{eq:lindbladMasterEquationDiagonal} of the Kitaev chain in Sec.~\ref{secThirdQuantizationKitaev}. We also prove that the master equation leads to thermalization at late times.

\subsection{Third Quantization} \label{secThirdQuantization}
We consider a general Lindblad master equation of the form
\begin{equation}
\dot{\rho}= \hat{\mathcal{L}}\rho = - i \left[H,\rho\right] + \sum_{\mu} \left(2L_{\mu}^{\phantom{\dagger}}\rho L_{\mu}^\dagger -\left\{L_{\mu}^\dagger L_{\mu}^{\phantom{\dagger}},\rho\right\}\right),
\label{eq:LindbladMasterEquationProsenForm}
\end{equation}
where the superoperator $\hat{\mathcal{L}}$ is called the Liouvillian. The Hamiltonian $H$ is assumed to be quadratic in some set of fermionic modes $d_m$ with $m=1,\dotsc,L$, and the jump operators $L_\mu$ are assumed to be linear. The formalism of Third Quantization then allows $\hat{\mathcal{L}}$ to be diagonalized, yielding an exact solution of the master equation. We outline this procedure for the case where $\hat{\mathcal{L}}$ is time-independent, see Ref.~\cite{ProsenZunkovic10} for details on the time-dependent case. The first step is to introduce $2L$ Hermitian Majorana operators $w_j$, written in terms of the $d_m$ (Dirac) fermions as
\begin{equation}
					w_{2m-1} = d_m^{\phantom{\dagger}} + d_m^{\,\dagger},\quad w_{2m} = i(d_m^{\phantom{\dagger}} - d_m^{\,\dagger}) , \quad m=1,\dotsc,L,
			\label{eq:majoranaCanonicalFermionsTransformationProsen}
			\end{equation}
and satisfying $\left\{w_j,w_k\right\}=2\delta_{jk}$. In terms of these Majorana operators, the Hamiltonian $H$ and jump operators $L_\mu$ in the master equation \eqref{eq:LindbladMasterEquationProsenForm} can be written as
			\begin{equation}
					H = \sum_{j,k=1}^{2L}{w_j H_{jk} w_k} = \ul{w}\cdot \mathbf{H} \ul{w}, \qquad L_\mu = \sum_{j=1}^{2L}{l_{\mu,j} w_j} = \ul{l}_\mu \cdot \ul{w},
			\label{eq:hamiltonianAndBathOperatorsProsen}
			\end{equation}
where $l_{\mu,j}$ are system--bath coupling constants and ${\ul{x}= [x_1,\ x_2,\ \ldots]^\text{T}}$ denotes a column vector. The $\mathbf{H}$ matrix can be chosen to be antisymmetric due to the anticommutation relations satisfied by the Majorana operators.

Next we define a $4^L$-dimensional vector space $\mathcal{K}$ of operators, spanned by the canonical basis $\ket{P_\ul{\alpha}}$ with 
			\begin{equation}
					P_{\alpha_1,\alpha_2,\ldots,\alpha_{2L}}= 2^{-L/2} w_1^{\alpha_1} w_2^{\alpha_2} \cdots w_{2L}^{\alpha_{2L}}, \quad \alpha_j \in \{0,1\},
			\end{equation} 
which forms an orthonormal basis with respect to the Hilbert-Schmidt inner product $\braket{x|y} = \Tr[x^\dagger y]$. Elements of $\mathcal{K}$ act on the fermionic Fock space $\mathcal{F}$, and the density matrix $\rho=\ket{\rho}$ is an element of $\mathcal{K}$. One can define on $\mathcal{K}$ a set of adjoint creation and annihilation superoperators according to
			\begin{equation}
					\hat{c}_j^\dagger \ket{P_\ul{\alpha}} = \delta_{\alpha_j,0} \ket{w_j P_\ul{\alpha}},\qquad \hat{c}_j \ket{P_\ul{\alpha}} = \delta_{\alpha_j,1} \ket{w_j P_\ul{\alpha}},
			\end{equation}
which satisfy the canonical anticommutation relations
			\begin{equation}
					\{\hat{c}_j,\hat{c}_k\} = 0, \quad \{\hat{c}_j,\hat{c}_k^\dagger\} = \delta_{jk}, \quad j,k = 1,\dotsc,2L.
			\label{eq:fermionicAntiCommutationRelations}
			\end{equation}
Note the use of hats to indicate superoperators acting on $\mathcal{K}$. In turn, this allows for the introduction of $4L$ Majorana superoperator maps 
			\begin{equation}
					\hat{a}_{2j-1}\coloneqq \frac{1}{\sqrt{2}}(\hat{c}_j^{\phantom{\dagger}} +\hat{c}_j^\dagger), \qquad \hat{a}_{2j}\coloneqq \frac{i}{\sqrt{2}}(\hat{c}_j^{\phantom{\dagger}} -\hat{c}_j^\dagger),
			\label{eq:HermitianMajoranaMaps}
			\end{equation}
which satisfy $\hat{a}_r^{\phantom{\dagger}} = \hat{a}_r^\dagger$ and the anticommutation relation $\{\hat{a}_r,\hat{a}_s\}=\delta_{rs}$. Using these Majorana maps, the master equation~\eqref{eq:LindbladMasterEquationProsenForm} governing the time evolution of $\rho=\ket{\rho}$ can be expressed in the Liouville--Fock picture as
			\begin{equation}
					\frac{d\ket{\rho}}{dt} = \hat{\mathcal{L}}\ket{\rho} = \left(\ul{\hat{a}}\cdot \mathbf{A}\ul{\hat{a}} - A_0 \hat{\mathbb{I}}\right) \ket{\rho}.
			\label{eq:liouvilleanInTermsOfStructureMatrix}
			\end{equation}
If we restrict our study to observables which are products of an even number of Majorana operators, then the antisymmetric structure matrix $\mathbf{A}$ of the Liouvillian has entries \cite{Prosen08} 
\begin{subequations}
			\begin{align}
					\mathbf{A}_{2j-1,2k-1} &= -2i \mathbf{H}_{jk} - \mathbf{M}_{kj} + \mathbf{M}_{jk},\label{e:A1}\\
					\mathbf{A}_{2j-1,2k} &= 2i\mathbf{M}_{kj},\\
					\mathbf{A}_{2j,2k-1} &= -2i\mathbf{M}_{jk},\\
					\mathbf{A}_{2j,2k} &= -2i \mathbf{H}_{jk} + \mathbf{M}_{kj} - \mathbf{M}_{jk},\label{e:A4}
			\end{align}
\end{subequations}
where
			\begin{equation}
					\mathbf{M}= \sum_\mu{\ul{l}_\mu \otimes \ul{l}_\mu^*}
			\label{eq:MmatrixDefinition}
			\end{equation}
is a $2L\times2L$ matrix encoding, via Eq.~\eqref{eq:hamiltonianAndBathOperatorsProsen}, details of the jump operators, while $A_0 = 2 \Tr[\mathbf{M}]$.

Assuming that the structure matrix $\mathbf{A}$ defined through Eqs.\ \eqref{e:A1}--\eqref{e:A4} can be diagonalized, there exist $4L$ eigenvectors $\ul{v}_q$, $q=1,\ldots,4L$ with corresponding (generally complex-valued) eigenvalues $r_1,-r_1,r_2,-r_2,\ldots,r_{2L},-r_{2L}$ where $\myRe(r_j)\geq0$. To be clear, $\ul{v}_{2n-1}$ has eigenvalue $r_n$, while the eigenvalue of $\ul{v}_{2n}$ is $-r_n$. The ordering is such that $\myRe(r_1)\geq \cdots \geq \myRe(r_{2L})$.  Hereafter the eigenvalues $r_j$ will be referred to as \emph{rapidities}. It can be shown that these rapidities coincide with the eigenvalues of the $2L \times 2L$ matrix \cite{KosProsen17}
\begin{equation}
\mathbf{X} = -2i \mathbf{H} + 2 \mathbf{M}_r,
\label{eq:XmatrixDefinitionProsen}
\end{equation}
where $\mathbf{H}$ is defined in Eq.~\eqref{eq:hamiltonianAndBathOperatorsProsen} and $\mathbf{M}_r=\myRe(\mathbf{M})$ is the real part of $\mathbf{M}$ in Eq.~\eqref{eq:MmatrixDefinition}. Using the $\ul{v}_q$ eigenvectors as rows, we construct the $4L\times4L$ matrix $\mathbf{V}$ and impose the normalization condition 
		\begin{equation}
					\mathbf{V}\mathbf{V}^\text{T}=\mathbb{I}_{2L} \otimes \sigma^x.
			\label{eq:normalisationConditionVMatrix}
			\end{equation}
It follows that $\mathbf{V}$ diagonalizes $\mathbf{A}$ such that $\mathbf{V}^{-\text{T}}\mathbf{A}\mathbf{V}^\text{T} = \mathbf{D}$ with $\mathbf{D}=\text{diag}\{r_1,-r_1,r_2,-r_2,\ldots,r_{2L},-r_{2L}\}$. Using the $\mathbf{V}$ matrix one can introduce a new set of superoperators in terms of which the Liouvillian $\hat{\mathcal{L}}$ is diagonal, and also calculate the steady state expectation value of any quadratic observable $w_j w_k$. For the latter application, we first introduce the notion of a nonequilibrium steady state $\ket{\text{NESS}}$, which satisfies $\hat{\mathcal{L}} \ket{\text{NESS}} = 0$ and hence is a stationary solution of the Lindblad equation \eqref{eq:liouvilleanInTermsOfStructureMatrix}. The nonequilibrium steady state is unique if, and only if, the rapidity spectrum $\{r_j\}$ does not contain zero. In this case the steady state expectation value of $w_j w_k$ is \cite{Prosen08}
			\begin{align}
					\langle w_j w_k\rangle_\text{NESS} =\text{ }& \delta_{jk} + \bra{\mathbb{I}}\hat{c}_j \hat{c}_k \ket{\text{NESS}} \nonumber\\
					=\text{ }&  \delta_{jk} + \frac{1}{2}\sum_{m=1}^{2L} (\mathbf{V}_{2m,2j-1}- i\mathbf{V}_{2m,2j})	\label{eq:expectationValuesMajoranasProsenSEC1}\\
					&\quad \quad \times (\mathbf{V}_{2m-1,2k-1}-i\mathbf{V}_{2m-1,2k}), \nonumber
			\end{align}
where $\ket{\mathbb{I}}$ represents the identity operator acting on the fermionic Fock space $\mathcal{F}$. The expectation values of more complicated observables containing even products of the fermionic operators can be determined using Wick contractions, see Ref.~\cite{Danielewicz84} for details.	

\subsection{Application to the open Kitaev chain} \label{secThirdQuantizationKitaev}
The master equation \eqref{eq:lindbladMasterEquationDiagonal} derived for the Kitaev chain in Sec.~\ref{s:OpenKitaev} shares the form of Eq.~\eqref{eq:LindbladMasterEquationProsenForm} and is therefore solvable through the Third Quantization procedure. To this end, the generic $d_m$ Dirac fermions will be identified with those of the chain's quasiparticle modes, in which case the Majorana fermions in Eq.~\eqref{eq:majoranaCanonicalFermionsTransformationProsen} become
	\begin{equation}
			w_{2n-1} = \eta_{k_n}^{\phantom{\dag}} + \eta_{k_n}^\dagger,\qquad w_{2n} = i\left(\eta_{k_n}^{\phantom{\dag}} - \eta_{k_n}^\dagger\right),
	\label{eq:majoranaCanonicalFermionsTransformationProsen5}
	\end{equation}
such that the diagonalized Kitaev chain Hamiltonian \eqref{eq:kitaevChainHamiltonianDiagonalForm} becomes
\begin{equation}
				H_\text{S} = \frac{i}{2}\sum_{n=1}^{L}{\lambda_{\phi,\alpha} (k_n) w_{2n}w_{2n-1}}
		\label{eq:kitaevChainDiagonalHamiltonianMajoranaFermions}
\end{equation}
with $k_n=2\pi n/L\ ({\rm mod}\ 2\pi)$. Similarly, the jump operators \eqref{eq:LindbladBathOperators} may be written as
\begin{equation}
L_{k_n,\pm} =  \frac{1}{2}\sqrt{\Gamma_{k_n,\pm}}\left(w_{2n-1}\pm iw_{2n}\right),
\label{eq:bathOp}
\end{equation}
with the bath coupling constants $\Gamma_{k_n,\pm}$ defined in Eqs.~\eqref{eq:LindbladBathCouplingConstants1} and \eqref{eq:LindbladBathCouplingConstants2}. The structures of $H_{\rm S}$ and $L_{k_n,\pm}$, in which $w_{2n}$ only couples to $w_{2n-1}$, ensure that the various quasiparticle modes evolve independently of each other, allowing the problem to be factorized into $L$ copies of a single fermion coupled to a bath. This is reflected in the structure matrix $\mathbf{A}$, which turns out to be block-diagonal, with a $4\times 4$ block associated to each quasiparticle mode. In the notation of Ref.~\cite{Prosen08} the block corresponding to the $\eta_{k_n}$ mode is
\begin{equation}
\mathbf{A}_n = -\frac{\lambda_{\phi,\alpha}(k_n)}{2} \mathbf{R} + \gamma \mathbf{B}_{\Gamma_{k_n,1},\Gamma_{k_n,2}},
\label{eq:aSubMatrices}
\end{equation}
with
\begin{equation}
\mathbf{R} \coloneqq  \begin{bmatrix}
	0&0&1&0\\
	0&0&0&1\\
	-1&0&0&0\\
	0&-1&0&0
\end{bmatrix}, 
\quad
\mathbf{B}_{x,y}\coloneqq  \begin{bmatrix}[1.3]
	0 & \frac{i}{2}x & -\frac{i}{2}y & \frac{1}{2}y\\
	-\frac{i}{2}x & 0 & \frac{1}{2}y & \frac{i}{2}y\\
	\frac{i}{2}y & -\frac{1}{2}y & 0 & \frac{i}{2}x\\
	-\frac{1}{2}y & -\frac{i}{2}y & -\frac{i}{2}x & 0
\end{bmatrix},
\label{eq:RandBmatrix}
\end{equation}
and $\Gamma_{k_n,1} = \Gamma_{k_n,+} + \Gamma_{k_n,-}$ and $\Gamma_{k_n,2} = \Gamma_{k_n,+} - \Gamma_{k_n,-}$. The eigenvalues of $\mathbf{A}_n$ then yield the two rapidities
\begin{equation}
		r_{\pm,n} = \tfrac{1}{2}\left[\gamma\Gamma_{k_n,1}\pm i\lambda_{\phi,\alpha}(k_n)\right].  
\label{eq:rapiditiesExpression}
\end{equation}
The $\mathbf{V}$ matrix, constructed using the eigenvectors of $\mathbf{A}$ as rows, exhibits a similar block-diagonal structure. The four eigenvectors of $\mathbf{A}_n$, paired according to their eigenvalues and normalized as per Eq.~\eqref{eq:normalisationConditionVMatrix}, yield the associated submatrix of $\mathbf{V}$ as
\begin{equation}
\mathbf{V}_n = \begin{bmatrix}[1.7]
	\frac{1}{2}\zeta_{n,-} \left(1+\frac{\Gamma_{k_n,2}}{\Gamma_{k_n,1}}\right) & -\frac{i}{2 \zeta_{n,-}} & \frac{i}{2}\zeta_{n,-} \left(1+\frac{\Gamma_{k_n,2}}{\Gamma_{k_n,1}}\right) & \frac{1}{2\zeta_{n,-}} \,\\
	\frac{1}{2\zeta_{n,-}} & \frac{i}{2\zeta_{n,-}} & -\frac{i}{2\zeta_{n,-}} & \frac{1}{2\zeta_{n,-}} \,\\
	-\frac{1}{2}\zeta_{n,+} \left(1-\frac{\Gamma_{k_n,2}}{\Gamma_{k_n,1}}\right) & \frac{i}{2 \zeta_{n,+}} & \frac{i}{2}\zeta_{n,+} \left(1-\frac{\Gamma_{k_n,2}}{\Gamma_{k_n,1}}\right) & \frac{1}{2\zeta_{n,+}} \,\\
	-\frac{1}{2\zeta_{n,+}} & -\frac{i}{2\zeta_{n,+}} & -\frac{i}{2\zeta_{n,+}} & \frac{1}{2\zeta_{n,+}}\,
\end{bmatrix}
\label{eq:vSubMatrices}
\end{equation}
with $\zeta_{n,\pm}=\sqrt{\Gamma_{k_n,1}/(2\Gamma_{k_n,\pm})}$.

\subsection{Thermalization of the open Kitaev chain}\label{secKitaevThermalization}
Except for a single mode precisely at one of the system's critical points, the rapidities in Eq.~\eqref{eq:rapiditiesExpression} are nonzero, which implies that the steady state of the Lindblad master equation for the Kitaev chain is unique. We now show that this steady state is a thermal state with a temperature matching that of the environment. In fact, the set of two-point correlations $\langle w_i w_j \rangle$ fully characterize this state \cite{Prosen08}. For the present example, these correlations are readily found by combining the general result in Eq.~\eqref{eq:expectationValuesMajoranasProsenSEC1} with the expression for $\mathbf{V}_n$ in Eq.~\eqref{eq:vSubMatrices}. The only nonzero off-diagonal correlations are found to be
\begin{equation}
	\langle w_{2n-1}w_{2n} \rangle=-\langle w_{2n}w_{2n-1} \rangle=i\frac{\Gamma_{k_n,2}}{\Gamma_{k_n,1}}
	\label{e:EtaMajoranaCorrelations}
\end{equation}
where, from Eqs.~\eqref{eq:LindbladBathCouplingConstants1} and \eqref{eq:LindbladBathCouplingConstants2}, one obtains
\begin{equation}
	\frac{\Gamma_{k_n,2}}{\Gamma_{k_n,1}} = -\tanh \left(\frac{ \lambda_{\phi,\alpha}(k_n)}{2T} \right).
	\label{eq:GammaRatioIdentity}
\end{equation}
Upon rewriting Eq.~\eqref{e:EtaMajoranaCorrelations} in terms of the operators for the quasiparticle modes via Eq.~\eqref{eq:majoranaCanonicalFermionsTransformationProsen5}, we find, as expected, that the steady-state occupation of these modes is given by a Fermi-Dirac distribution,
\begin{equation}
	\langle \eta_{k_n}^\dagger \eta_{k_n}^{\phantom{\dag}} \rangle = \frac{1}{2}\big(1 + i \langle w_{2n} w_{2n-1} \rangle\big) = \left(e^{\lambda_{\phi,\alpha}(k_n)/T}+1\right)^{-1}. 
\end{equation}
This confirms that the coupling of each lattice site to an individual bath, as set out in Sec.~\ref{s:OpenKitaev}, indeed thermalizes the system.


\section{Dynamics of two-point correlations and mode occupation numbers} \label{s:secTimeEvolutionCorrelations}
The Third Quantization formalism described in Sec.~\ref{secThirdQuantization} also provides a convenient setting for studying the dynamics of the correlations between fermionic variables \cite{KosProsen17}. In the following two subsections we outline this approach and show how it can be applied to the Kitaev chain to study the time evolution of mode occupation numbers during various types of time-dependent protocols. 

\subsection{Correlation dynamics} \label{secCorrelationDynamics}
We combine all two-point correlations of the $w_i$ Majorana fermions into the correlation matrix $\mathbf{C}$ with elements
\begin{equation}
\mathbf{C}_{jk} = \Tr[w_j w_k \rho] - \delta_{jk} = \braket{\mathbb{I}|\hat{c}_{j} \hat{c}_{k}|\rho}. 
\label{eq:correlationMatrixEntriesMainText}
\end{equation}
The dynamics of $\mathbf{C}$ can be derived from the Liouville--Fock master equation \eqref{eq:liouvilleanInTermsOfStructureMatrix} and is governed by
\begin{equation}
-\frac{1}{2} \frac{\text{d}\mathbf{C}(t)}{\text{d}t} = \mathbf{X}^\text{T} \mathbf{C}(t) + \mathbf{C}(t)\mathbf{X} + 4i \mathbf{M}_i
\label{eq:matrixDiffEquationWithXandM}
\end{equation}
with $\mathbf{X} = -2i \mathbf{H} + 2 \mathbf{M}_r$ as introduced in Eq.~\eqref{eq:XmatrixDefinitionProsen}, and $\mathbf{M}_i$ the imaginary part of $\mathbf{M}$ in Eq.~\eqref{eq:MmatrixDefinition}.  See Ref.~\cite{KosProsen17} for a derivation of this result. From $\mathbf{C}(t)$, higher-order correlations can found through Wick contractions.

\subsection{Mode occupation numbers of the Kitaev chain} \label{secOccupationNumbersKitaev}
Turning to the Kitaev chain, we recall that the Majorana fermions $w_j$ appearing in Eq.~\eqref{eq:correlationMatrixEntriesMainText} are defined in terms of Dirac fermions $d_m$ through Eq.~\eqref{eq:majoranaCanonicalFermionsTransformationProsen}. It is, however, a matter of choice whether one uses the Dirac site fermions $c_m$ of the Kitaev chain for the definition of the Majorana fermions, or the Dirac quasiparticles $\eta_m$, or other linearly related choices. When investigating the steady state of the open Kitaev chain in Sec.~\ref{s:diagonaliseLindbladian}, we identified the $d_m$ fermions with those of the model's quasiparticle modes. However, this choice would introduce complications when considering protocols with time-dependent parameters in the Hamiltonian. Such time-dependencies generally affect how the quasiparticle modes are defined, introducing an additional time-dependence in the correlation matrix \eqref{eq:correlationMatrixEntriesMainText} through the Majorana operators themselves. In turn, this would generate new terms in the equation of motion \eqref{eq:matrixDiffEquationWithXandM}. While this scenario, as shown in Sec.~\ref{s:example}, can provide useful insights, for a direct numerical approach it is preferable to work in terms of $d_m$ fermions that are unaffected by parameter ramps. Here a convenient choice is to use the $a_k$ Fourier fermions introduced in Sec.~\ref{ss:Kitaev Diagonalization}. This choice also brings about a factorization of the full equation of motion for $\mathbf{C}$ in Eq.~\eqref{eq:matrixDiffEquationWithXandM} into a set of lower-dimensional equations. This is seen from the forms of the Hamiltonian $H_{\rm S}$ \eqref{eq:lrkMatrixForm} and jump operators \eqref{eq:LindbladBathOperators} in which the $a_k$ Fourier mode only couples to its time-reversed partner $a_{-k}$. This imparts a block-diagonal structure to the $\mathbf{X}$ and $\mathbf{M}$ matrices in Eq.~\eqref{eq:matrixDiffEquationWithXandM}, with a $4\times 4$ block for each $\{+k,-k\}$ pair of Fourier modes, and $2\times 2$ blocks for the $k=0$ and, if present, $k=\pi$ modes. In particular, the Lindblad dynamics captured by Eq.~\eqref{eq:matrixDiffEquationWithXandM} cannot modify the correlations between Fourier modes which are not time-reversed pairs. If only correlations between $\{+k,-k\}$ mode pairs are present initially, as would be the case if $\rho(0)$ is a thermal state, then this will remain the case at all later times. 

We first consider the correlations associated with a particular $\{+k,-k\}$ pair of Fourier modes, keeping in mind that we are ultimately interested in the occupations of the two associated quasiparticle modes. The relevant Majorana operators are
\begin{subequations}
		\begin{align}
		w_{1} &= a_{k}^{\phantom{\dag}} + a_{k}^\dagger,& w_{2} &= i(a_{k}^{\phantom{\dag}} - a_{k}^\dagger),\label{eq:FourierMajoranasDefinition1}\\
		w_{3} &= a_{-k}^{\phantom{\dag}} + a_{-k}^\dagger,& w_{4} &= i(a_{-k}^{\phantom{\dag}} - a_{-k}^\dagger),\label{eq:FourierMajoranasDefinition2}
	\end{align}
\end{subequations}
and the associated $4\times4$ correlation matrix reads
\begin{equation}
[\mathbf{C}_{\{k,-k\}}]_{ij} = \Tr[w_i w_j \rho]-\delta_{ij}, \quad i,j=1,\ldots,4.
\label{eq:CkMinuskModeDefinition}
\end{equation}
From Eq.~\eqref{eq:matrixDiffEquationWithXandM} it follows that the dynamics of $\mathbf{C}_{\{k,-k\}}$ is governed by the $4\times 4$ matrix equation
\begin{multline}
-\frac{1}{2} \frac{\text{d}\mathbf{C}_{\{k,-k\}}(t)}{\text{d}t} = \mathbf{X}_{\{k,-k\}}^\text{T} \mathbf{C}_{\{k,-k\}}(t) \label{eq:matrixDiffEquationWithXandM4by4}\\
+ \mathbf{C}_{\{k,-k\}}(t)\mathbf{X}_{\{k,-k\}} + 4i \myIm(\mathbf{M}_{\{k,-k\}})
\end{multline}
with
\begin{equation}
		\mathbf{X}_{\{k,-k\}} = -2 i \mathbf{H}_{\{k,-k\}} + 2 \myRe(\mathbf{M}_{\{k,-k\}}).
	\label{eq:XmatrixDefinition4by4}
	\end{equation}
The matrix $\mathbf{H}_{\{k,-k\}}$, provided  in Appendix~\ref{app:HandMmatrices}, is obtained from Eq.~\eqref{eq:hamiltonianAndBathOperatorsProsen} after rewriting the Kitaev Hamiltonian \eqref{eq:lrkMatrixForm} in terms of the four Majorana operators introduced in Eqs.\ \eqref{eq:FourierMajoranasDefinition1} and \eqref{eq:FourierMajoranasDefinition2}. To determine $\mathbf{M}_{\{k,-k\}}$ we express the jump operators \eqref{eq:LindbladBathOperators} in terms of the Majorana operators via the Bogoliubov transformation in Eq.~\eqref{eq:bogo}, and then combine definitions \eqref{eq:hamiltonianAndBathOperatorsProsen} and \eqref{eq:MmatrixDefinition} to arrive at the final expression \eqref{eq:MmatrixDefinition4by4}. The two-point correlations solved for from Eq.~\eqref{eq:matrixDiffEquationWithXandM4by4} also give access to the dynamics of the excitation density, i.e., the occupation numbers of the $\eta_k$ modes. The latter will be the main quantity of interest in the example discussed in Sec.~\ref{s:example}.
	
The set of lower-dimensional equations in Eq.~\eqref{eq:matrixDiffEquationWithXandM4by4} can be further simplified by using Eqs.~\eqref{eq:FourierMajoranasDefinition1} and \eqref{eq:FourierMajoranasDefinition2} to write down equations of motions for the correlations of Dirac Fourier fermions,
\begin{widetext}
\begin{subequations}
\begin{align}
	\frac{\text{d}}{\text{d}t}\ev{a_k^\dag a_k^{\phantom\dag}} = &\gamma\left(\Gamma_{k,1}+\Gamma_{k,2}\cos(2\beta_k)-2\Gamma_{k,1}\ev{a_k^\dag a_k^{\phantom\dag}}\right)-\lambda_k\sin(2\beta_k)\left(\ev{a_k^\dag a_{-k}^\dag}^*+\ev{a_k^\dag a_{-k}^\dag}\right),\label{e:FourierFermionCorrelations1}\\
	\frac{\text{d}}{\text{d}t}\ev{a_k^\dag a^\dag_{-k}} = &\gamma\left(i\Gamma_{k,2}\sin(2\beta_k)-2\Gamma_{k,1}\ev{a_k^\dag a^\dag_{-k}}\right)+\lambda_k\left[\sin(2\beta_k)\left(2\ev{a_k^\dag a_{k}^{\phantom\dag}}-1\right)+2i\cos(2\beta_k)\ev{a_k^\dag a_{-k}^\dag}\right].\label{e:FourierFermionCorrelations2}
\end{align}
\end{subequations}
\end{widetext}
Using the Bogoliubov transformation \eqref{eq:bogo}, the solutions to these equations also provide access to the mode occupation numbers
\begin{equation}
	\ev{\eta_k^\dag\eta_k^{\phantom\dag}}=\cos(2\beta_k)\left(\ev{a_k^\dag a_k^{\phantom\dag}}-\tfrac{1}{2}\right)+\sin(2\beta_k)\myIm[\ev{a_k^\dag a_{-k}^\dag}]+\tfrac{1}{2}.
\label{e:OccupationFromFourierFermionCorrelations}
\end{equation}
	

\section{Example: Parameter ramp at nonzero temperature} \label{s:example}
We now turn to applications that illustrate the usefulness of our results for probing the nonequilibrium dynamics of the open Kitaev chain. In particular, we study the excitation dynamics resulting from ramping protocols in which a parameter in the Kitaev chain Hamiltonian is driven towards, or across, its quantum critical value. This offers the opportunity to explore the emergence of Kibble-Zurek physics and universal scaling laws in the context of an open quantum system. In a first scenario, we fix the parameters in the system Hamiltonian at their critical values and then ramp the bath temperature from a positive initial value down to zero. We explore this protocol for cooling the system to criticality, and the scaling laws emerging from it, in our companion paper \cite{KingKastnerKrielLetter}. Here we focus instead on ramps of a parameter in the Hamiltonian, performed at a fixed positive temperature, where the parameter crosses its critical value. The zero-temperature version of this protocol has been studied previously for the Kitaev chain \cite{DuttaDutta17}, and also for a range of other systems exhibiting quantum phase transitions \cite{Dziarmaga10}. Central to these studies are the scaling laws obtained for the excitation density at the end of the ramp, which can be understood within the framework of the Kibble-Zurek mechanism as a consequence of the breakdown of adiabatic evolution at the critical point. Although quantum phase transitions are strictly zero temperature phenomena, it is expected that a system at small nonzero temperatures will be sensitive to the proximity of the critical point, and that in certain parameter regimes universal scaling behavior might still emerge. This question was investigated using Keldysh techniques in Refs.~\cite{Patane_etal08,Patane_etal09} for a spin model, but with a different form of system--bath coupling. More recently, the authors of Ref.~\cite{RossiniVicari20} considered a class of open systems governed by Lindblad master equations with local jump operators and argued that Kibble-Zurek scaling can persist in the presence of weak dissipation. Here, too, the Kitaev chain was used for numeric benchmarking. In contrast, our open-system model has nonlocal jump operators, which are essential for enforcing thermalization at late times, which in turn allows for a well-defined temperature that can be associated with the environment.

The specific example we consider is a Kitaev chain initialized in a thermal state with temperature $T$, which matches the temperature of the baths. The bath spectral density $\mathcal{J}$, which appears in the bath coupling constants~\eqref{eq:LindbladBathCouplingConstants1} and \eqref{eq:LindbladBathCouplingConstants2}, is arbitrary at this point. In our numeric implementation we will specialize to an ohmic density $\mathcal{J}(\lambda)=\pi\delta\lambda \exp(-\lambda/\lambda_c)$
with dimensionless parameter $\delta$ and a high bath cutoff frequency $\lambda_c\gg \lambda_k$. The chemical potential $\mu$ in the Kitaev chain Hamiltonian \eqref{eq:LRKitaevChainHamiltonian} is then ramped across its critical value $\mu_c$ while keeping the bath temperature fixed (see Fig.~1 of Ref.~\cite{KingKastnerKrielLetter}). We consider the $k=0$ critical point, as identified in Sec.~\ref{ss:Kitaev Quantum Phases}. The ramp is performed linearly, with $\mu(t)$ increasing from $\mu_i$ to $\mu_f$ at a constant velocity $v$, such that
\begin{equation}
\mu(t) = v t +\mu_i, \quad t\in[0,t_f = (\mu_f-\mu_i)/v].
\label{eq:muRampDefinitionChapter5}
\end{equation}
Solving for the correlations from Eqs.~\eqref{e:FourierFermionCorrelations1} and \eqref{e:FourierFermionCorrelations2} numerically, and then using Eq.~\eqref{e:OccupationFromFourierFermionCorrelations}, allows us to monitor the system's excitation density 
\begin{equation}\label{eq:excitationdensity}
\mathcal{E}(t)=\frac{1}{L}\sum_k \braket{\eta_k^\dagger \eta_k^{\phantom{\dagger}}}(t)
\end{equation}
throughout the ramp. There are two physical mechanisms which impact on the dynamics of $\mathcal{E}(t)$. The first are the coherent excitations, which result from the breakdown of the adiabaticity of the system's unitary dynamics close to the critical point. This is an inevitable consequence of the critical slowing down of the system dynamics due to the closing of the excitation gap at the critical point. The second mechanism is that of incoherent thermal excitations, and possible dissipation, resulting from the system--bath coupling. As emphasized in Refs.~\cite{Patane_etal08,Patane_etal09} these processes cannot generally be disentangled to write $\mathcal{E}(t)$ as a sum of two independent contributions, except in limiting cases. Instead, there is a complicated interplay between the two mechanisms. Despite this, it is possible to cast the compact equations~\eqref{e:FourierFermionCorrelations1} and \eqref{e:FourierFermionCorrelations2} into a form with a somewhat more transparent structure. To this end, we invert the Bogoliubov transformation in Eq.~\eqref{eq:bogo} and reformulate these evolution equations as ones describing the dynamics of $\eta_k$ correlations. Here it is crucial to account for the time-dependence that the ramping protocol introduces into the Bogoliubov angle $\beta_k(t)$ via the latter's dependence on $\mu(t)$. The resulting equations are
\begin{subequations}
\begin{align}
	\frac{\text{d}}{\text{d}t}\ev{\eta_k^\dag \eta_k^{\phantom\dag}} = & -\gamma\Gamma_{k,1}\left(2\ev{\eta_k^\dag \eta_k^{\phantom\dag}}-\frac{\Gamma_{k,2}}{\Gamma_{k,1}}-1\right)\nonumber\\
	&+i\dot{\beta}_k\left[\ev{\eta^\dag_k\eta^\dag_{-k}}^*-\ev{\eta^\dag_k\eta^\dag_{-k}}\right],\label{e:EtaFermionCorrelations1}\\
	\frac{\text{d}}{\text{d}t}\ev{\eta_k^\dag \eta^\dag_{-k}} = &-2\gamma\Gamma_{k,1}\ev{\eta_k^\dag \eta^\dag_{-k}}-2i\dot{\beta}_k\left(\ev{\eta_k^\dag \eta_k^{\phantom\dag}}-\tfrac{1}{2}\right)\nonumber\\
	&+2i\lambda\ev{\eta_k^\dag \eta^\dag_{-k}}\label{e:EtaFermionCorrelations2}.
\end{align}
\end{subequations}
Let us unpack how the coherent and incoherent excitation mechanisms are contained in these equations. Setting $\gamma=0$ would eliminate the coupling to the bath and isolate the coherent excitations. The resulting dynamics is then equivalent to that of the standard two-level Landau--Zener problem. In this case the rate at which excitations are generated, i.e., the right hand side of Eq.~\eqref{e:EtaFermionCorrelations1}, is proportional to $\dot{\beta}_k$, which is itself a measure of how rapidly the chain's Hamiltonian varies during the ramp. As suggested by the adiabatic theorem, larger $\dot{\beta}_k$ will lead to greater deviations from adiabatic dynamics, and a more rapid generation of excitations. In contrast, retaining only the term proportional to $\gamma$ in Eq.~\eqref{e:EtaFermionCorrelations1} isolates the bath-induced dynamics. In this case the expression for $\Gamma_{k,2}/\Gamma_{k,1}$ in \eqref{eq:GammaRatioIdentity} allows us to write
\begin{equation}
	\frac{\text{d}}{\text{d}t}\ev{\eta_k^\dag \eta_k^{\phantom\dag}} = -2\gamma\Gamma_{k,1}\left[\ev{\eta_k^\dag \eta_k^{\phantom\dag}}-n(\lambda_k/T)\right]
	\label{eq:RateEquation}
\end{equation}
with $n(\lambda_k/T)=(\exp(n(\lambda_k/T))+1)^{-1}$ the Fermi-Dirac distribution. This rate equation describes the relaxation of the mode occupation to its thermal equilibrium value. The associated relaxation rate of the mode is the prefactor $r_k=2\gamma\Gamma_{k,1}$. Note that varying the bath temperature while keeping the system Hamiltonian fixed would imply that $\dot{\beta_k}=0$, in which case Eq.~\eqref{eq:RateEquation} provides a full account of the excitation dynamics. This rate equation will serve as a starting point for analyzing the scaling laws resulting from temperature ramps at the critical point in our companion paper \cite{KingKastnerKrielLetter}. For the present example, we turn to numeric results in order to illustrate the interplay between the two mechanisms highlighted above.

\begin{figure}[t]%
\includegraphics[width=\columnwidth]{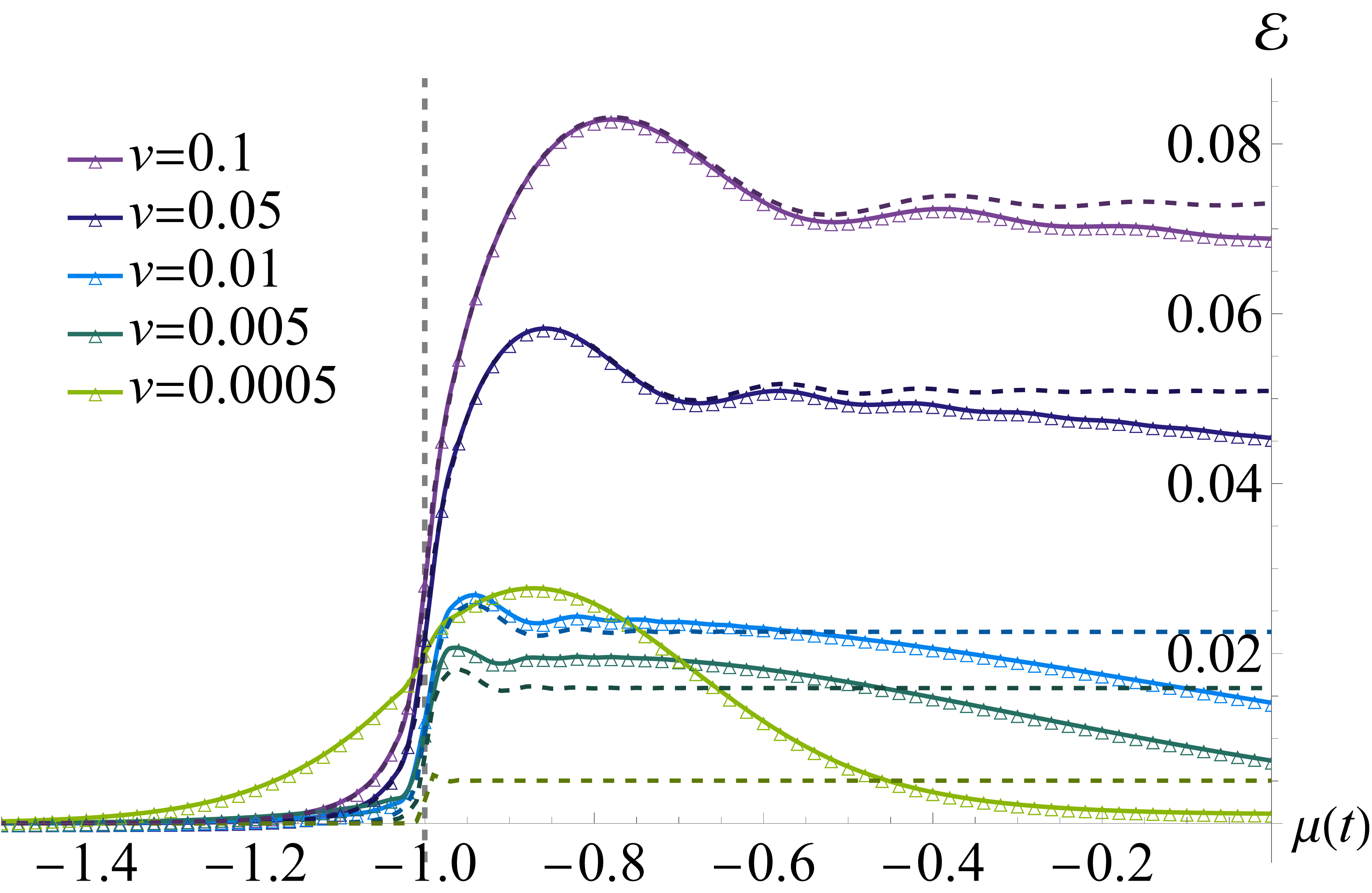}%
\caption{The excitation density $\mathcal{E}$ \eqref{eq:excitationdensity} of the nearest-neighbor Kitaev chain as a function of $\mu(t)$ for finite-temperature ramps with various velocities, calculated by solving Eqs.~\eqref{e:FourierFermionCorrelations1} and \eqref{e:FourierFermionCorrelations2}. Dashed lines in the corresponding colors show the excitation density for purely coherent excitations, i.e., $\mathcal{E}$ when $\gamma=0$. The vertical dashed line marks the chemical potential's critical value $\mu_c=-1$.
Parameter values for the computations are $L=2^{12}$, $J=\Delta=1$, $\mu_i=-5$, $\mu_f=0$, $T=0.181$, $\delta=1$, $\lambda_c=4000$ and $\gamma=0.001$.}%
\label{fig:evolution}%
\end{figure}

\begin{figure}[t]%
\includegraphics[width=\columnwidth]{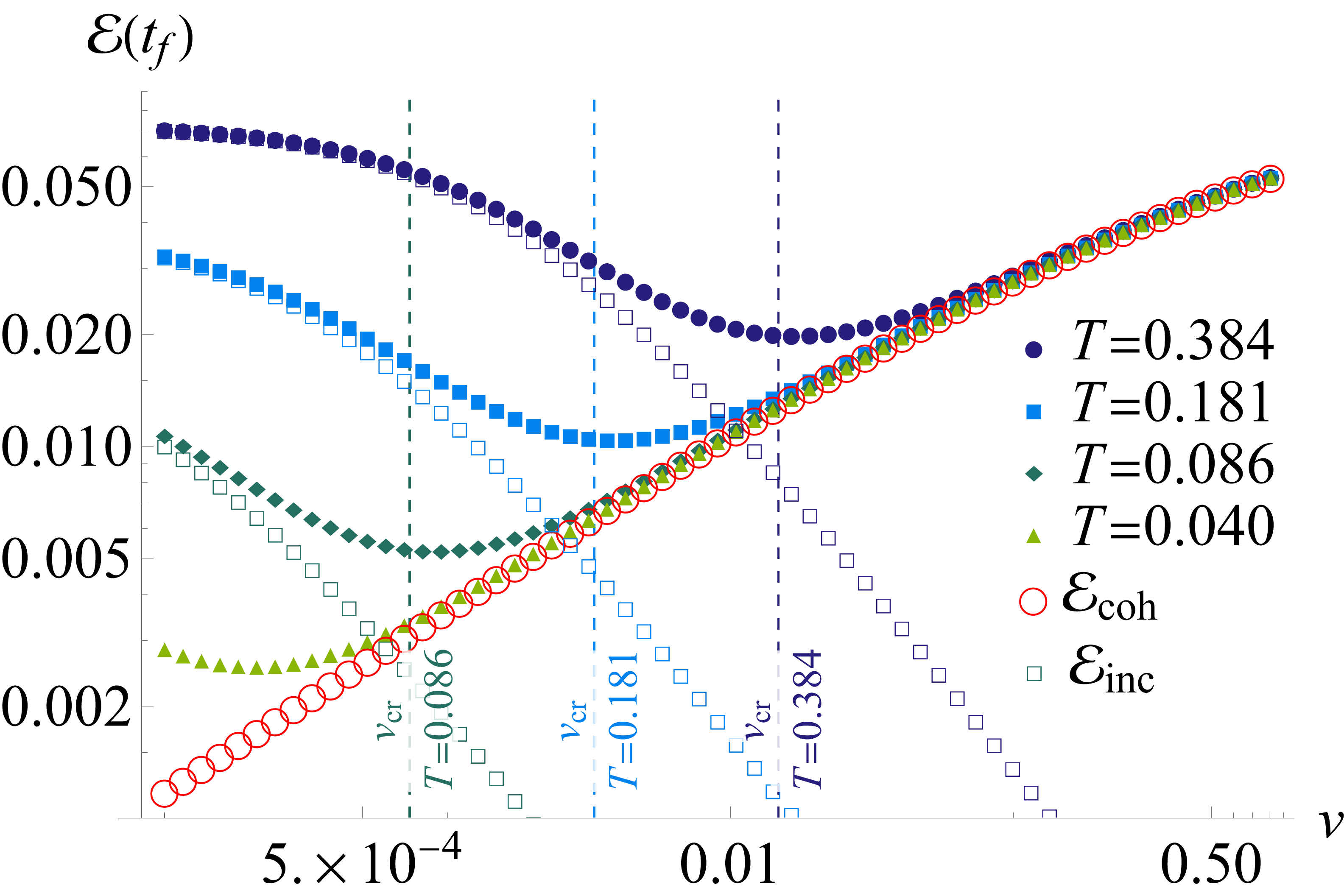}%
\caption{The total excitation density $\mathcal{E}(t_f)$ at the end of the ramping protocol as a function of the velocity $v$ for various fixed finite temperatures. Dashed vertical lines indicate quantitative crossover velocities $v_{\text{cr}}$ that separate the regimes in which coherent and incoherent mechanisms are dominant. The coherent and incoherent contributions to the excitation density are also shown separately as empty circle and square data points, respectively.   
Parameter values for these computations are $L=2^{12}$, $J=\Delta=1$, $\mu_i=-3$, $\mu_f=-1$, $\delta=1$, $\lambda_c=4000$ and $\gamma=0.001$.}%
\label{fig:totalEDvsV}%
\end{figure}

We present two sets of numeric results for the excitation density \eqref{eq:excitationdensity}, obtained through an exact numeric treatment of the coupled differential equations \eqref{e:FourierFermionCorrelations1} and  \eqref{e:FourierFermionCorrelations2} using the Bulirsch--Stoer method with a semi-implicit extrapolation modification due to Bader and Deuflhard \cite{Press_etal92}. This numeric approach was necessitated by the high degree of stiffness exhibited by these equations. For the nearest-neighbor Kitaev chain, Fig.~\ref{fig:evolution} shows the time evolution of the excitation density during the ramp, and includes for comparison results for the closed chain as well. For all but the slowest ramp velocity we observe a rapid increase in the excitation density as adiabatic evolution breaks down in the vicinity of the critical point at $\mu=\mu_c=-1$. However, in contrast to the case of an isolated chain with only coherent excitations (dashed curves in Fig.~\ref{fig:evolution}), we observe (i) the generation of thermal excitations close to $\mu_c$, particularly for slow velocities, and (ii) bath-induced relaxation after the critical point is crossed. The former suggests an interesting interplay between the two mechanisms of excitation, with the ramp velocity being a determining factor in which of the two is dominant. The second observation is a consequence of the system leaving the quantum critical region and crossing into the region where $T<\Delta$ \cite{Patane_etal09}. To better illustrate observation (i), we show in Fig.~\ref{fig:totalEDvsV} the final excitation density $\mathcal{E}(t_f)$ as a function of the ramp velocity for a ramp terminating at $\mu_f=\mu_c$. Here it is possible to identify a quantitative crossover velocity $v_\text{cr}$ which separates the $v < v_\text{cr}$ region where incoherent bath-induced excitations are dominant from the $v > v_\text{cr}$ region where coherent excitations will dominate. The range of ramp velocities for which each mechanism is dominant is crucial for understanding why scaling behavior was absent in previous studies \cite{Patane_etal08,Patane_etal09,RossiniVicari20} for certain regimes of parameter space.


\section{Summary and Conclusions}
\label{s:summary}

The objective of the present paper was to construct and solve a ``minimal model'' with which nonequilibrium phenomena in many-body open quantum systems can be studied analytically under time-dependent parameter changes in the system and/or the bath. In particular, the following properties were asked for:
\makeatletter
\let\orig@listi\@listi
\def\@listi{\orig@listi\topsep=0\baselineskip}
\makeatother
\begin{enumerate}
\setlength{\itemsep}{0pt}
\setlength{\parskip}{0pt}
\item The master equation modeling the open-system dynamics is exactly solvable, such that large systems can be dealt with.\label{i:1}
\item The open system thermalizes at late times under the time evolution induced by the master equation,\label{i:2}
\item The corresponding closed system undergoes an equilibrium quantum phase transition, ideally accompanied by further interesting physical properties, like topological order, whose fate in nonequilibrium conditions and under the influence of a thermal environment can then be studied.\label{i:3}
\end{enumerate}
Our model of choice is a Kitaev chain to which identical baths are coupled individually at each lattice site. We demonstrate the working of a selection of methods which are well suited to this model and these types of studies, and which open up avenues for a range of applications to be explored. The quadratic nature of the Kitaev chain, in combination with linear system--bath couplings, allows for a self-consistent derivation of the corresponding Lindblad master equation by a method due to D'Abbruzzo and Rossini \cite{DAbbruzzoRossini21}, resulting in a particularly reliable and faithful description of the open-system dynamics. The resulting Lindbladian \eqref{eq:lindbladMasterEquationDiagonal}--\eqref{eq:LindbladBathCouplingConstants2} is quadratic in the fermionic operators and can hence be diagonalized in Liouville space by the method of Third Quantization \cite{Prosen08}. Suitable model-specific adaptations of this method result in an analytic solution of a relatively simple form. Specifically, we obtain a set of decoupled $4\times4$ matrix differential equations \eqref{eq:matrixDiffEquationWithXandM4by4}, which in turn give access to the dynamics of arbitrary correlation functions and mode occupation numbers of the open Kitaev chain. For nonequilibrium protocols with time-dependent Hamiltonian or bath parameters, these differential equations have time-dependent coefficients, but numerical integration can nonetheless give access to systems consisting of thousands of lattice sites.

A broad variety of nonequilibrium situations and protocols of quantum many-body systems at finite temperature can be studied with the methods proposed and discussed in this paper. This includes protocols in which system parameters are quenched or ramped at fixed finite temperature, like in the example discussed in Sec.~\ref{s:example}, or ramps of the bath temperature, a situation that is explored in detail in the companion paper \cite{KingKastnerKrielLetter}. Further generalizations include protocols in which a Hamiltonian parameter and the bath temperature are ramped simultaneously, as well as nonlinear ramps, all of which we are planning to explore in future work. Monitoring excitation densities during such ramps can further the understanding of the interplay between the coherent excitations which result from a breakdown in adiabaticity, and the bath-induced heating and cooling of the system. A similar interplay of coherent and dissipative effects, but in a rather different setting, may be studied by considering oscillating Hamiltonian and/or bath parameters, allowing for the investigation of Floquet physics at finite temperature.

The derivations in this paper, and also the applications just mentioned, lead to  expressions of moderate complexity, partly because of translation invariance of system and bath (or, in an open-system description, translation invariance of the corresponding master equation). Several interesting extensions of our results become possible once translation invariance is abandoned. This could be done by introducing disorder into the hopping or pairing strengths of the Kitaev chain, resulting in a setting in which the fate of many-body localization in the presence of a thermal bath can be investigated. Another way of breaking translation invariance is by introducing a non-uniform set of thermal baths, with different lattice sites coupled to baths at different temperatures. Even in the absence of explicitly time-dependent parameters or temperatures, such a setting gives rise to nonequilibrium behavior, characterized by nonequilibrium steady states. Augmenting this setting with a time-dependent parameter in the Hamiltonian,  universal scaling features may be discovered when the Hamiltonian is parametrically driven across its critical value. To the best of our knowledge, the scaling behavior of excitations on top of nonequilibrium steady states is an unexplored field at present, which may greatly benefit from the concepts and methods put forward and advertised in this paper.

\acknowledgments
E.C.K.\ acknowledges financial support by the National Institute for Theoretical Physics of South Africa through an MSc bursary.


\appendix 

\section{The Lamb-shift Hamiltonian} \label{app:LSHamiltonian}
The Lamb-shift Hamiltonian entering in Eq.~\eqref{eq:lindbladMasterEquationDiagonal} can be expressed in terms of the eigenoperators $O_n(\lambda)$ \eqref{eq:eigenoperatorsGeneral} as \cite{DAbbruzzoRossini21,BreuerPetruccione}
\begin{equation}
H_\text{LS} = \gamma\sum_\lambda \sum_{m,n}{S_{m n}(\lambda) O_m^\dagger(\lambda) O_n(\lambda)},
\label{eq:HLS}
\end{equation}
where
\begin{align}\label{eq:Sdefinition}
S_{mn} (\lambda) &= \frac{1}{2i}\int_0^\infty d\tau \, \left[e^{i\lambda \tau} \langle \tilde{R}^\dagger_m(\tau)R_n \rangle-e^{-i\lambda \tau} \langle R^\dag_m \tilde{R}_n(\tau) \rangle\right].
\end{align}
As was done in Eq.~\eqref{eq:lindbladMasterEquationDiagonal}, we have introduced $\gamma$ as a measure of the system--bath coupling strength, which is taken to be weak. The above expressions can be simplified through the same series of steps as was used for the dissipator in Sec.~\ref{sec:dissipator}, yielding $S_{mn} (\lambda)=\delta_{mn}S(\lambda)$ and
\begin{equation}
	H_\text{LS}=\gamma\sum_k\left[S(\lambda_k)-S(-\lambda_k)\right]\eta^\dag_k \eta_k^{\phantom{\dag}}.
	\label{eq:LShamiltonianKitaevChain}
\end{equation}


\section{Further restrictions on the \texorpdfstring{$\bfvec{T}$}{T} transformation} \label{app:symmetries}

Introducing the column vector $\Phi=[c_1,\ldots,c_{L},c_1^\dagger,\ldots,c_{L}^\dagger]^\text{T}$, we can write the Kitaev chain Hamiltonian \eqref{eq:LRKitaevChainHamiltonian} as
\begin{equation}
	H_{\rm S}=\frac{1}{2}\Phi^\dag \mathbf{H}_{\rm BdG}\Phi,
\end{equation}
where $\mathbf{H}_{\rm BdG}$ is a real and symmetric $2L\times 2L$ matrix. The transformation $\mathbf{T}$ in Eq.~\eqref{eq:canonicalTransformation} then diagonalizes $\mathbf{H}_{\rm BdG}$ as
\begin{equation}
	\mathbf{T}^\dag \mathbf{H}_\text{BdG}\mathbf{T}=\mat{\mathbf{\Lambda} & 0}{0 & -\mathbf{\Lambda}},
\end{equation}
where $\mathbf{\Lambda}$ is an $L\times L$ diagonal matrix containing the mode energies $\lambda_k$. To match the case discussed in Sec.~\ref{sec:dissipator} of the main text, we assume that these energies are positive. If $\bfvec{v}$ is a positive-energy eigenvector of $\mathbf{H}_{\rm BdG}$, then, since the latter is real, so too is $\bfvec{v}^*$. In particular, $\bfvec{v}^*$ will be orthogonal to any negative-energy eigenvector of $\mathbf{H}_{\rm BdG}$. Applying this reasoning to the eigenvectors forming the columns of $\mathbf{T}$, we conclude that 
\begin{equation}
	\cvec{\mathbf{V}^*}{\mathbf{S}}^\dag\cvec{\mathbf{S}}{\mathbf{V}^*}=\mathbf{V}^\text{T}\mathbf{S}+\mathbf{S}^\dag \mathbf{V}^*=\mathbf{0},
\end{equation}
which is the additional restriction on $\mathbf{T}$ used in the derivation of Eq.~\eqref{eq:phiUnitaryCondition}.


\section{Explicit forms of the matrices in Eq.\texorpdfstring{~\eqref{eq:matrixDiffEquationWithXandM4by4}}{ (63)}.} \label{app:HandMmatrices}
The Hamiltonian and Lindblad matrices appearing in the dimensionally-reduced version of the equation of motion \eqref{eq:matrixDiffEquationWithXandM4by4} for the correlation matrix are
\begin{widetext}
\vspace{-0.28cm}
	\begin{align}
		\mathbf{H}_{\{k,-k\}} &= \frac{i}{4} \lambda_{\phi,\alpha}(k) \begin{bmatrix} 
			0 &   - \cos(2 \beta_{k}) & - \sin(2 \beta_{k}) & 0\\
			\cos(2 \beta_{k}) & 0 & 0 & \sin(2 \beta_{k})   \\
			\sin(2 \beta_{k})  & 0 & 0 &  -\cos(2 \beta_{k})\\
			0 & -\sin(2 \beta_{k})  & \cos(2 \beta_{k}) & 0
		\end{bmatrix},	\label{eq:Hsubkminusk}\\
		\mbox{}\nonumber\\
		\mathbf{M}_{\{k,-k\}} &= \frac{\gamma}{4} \begin{bmatrix}
			\Gamma_{k,1} &  -i\Gamma_{k,2}\cos(2 \beta_{k}) & -i \Gamma_{k,2} \sin(2 \beta_{k}) & 0\\
			 i\Gamma_{k,2}\cos(2 \beta_{k}) & \Gamma_{k,1} & 0 &  i \Gamma_{k,2} \sin(2 \beta_{k})  \\
			 i \Gamma_{k,2} \sin(2 \beta_{k})  & 0 & \Gamma_{k,1} &  -i \Gamma_{k,2}\cos(2 \beta_{k})\\
			0 & - i \Gamma_{k,2} \sin(2 \beta_{k}) & i \Gamma_{k,2} \cos(2 \beta_{k}) & \Gamma_{k,1}
		\end{bmatrix},	\label{eq:MmatrixDefinition4by4}\\
		\mbox{}\nonumber\\
		\mathbf{H}_{k=0,\pi} &= \frac{i}{4} \lambda_{\phi,\alpha}(k) (1-2\delta_{k0}) \begin{bmatrix}
			0 &  1\\
			-1 & 0
		\end{bmatrix},\qquad\qquad \mathbf{M}_{k=0,\pi} = \frac{\gamma}{4}\mat{\Gamma_{k,1} &  i \Gamma_{k,2}}{
			-i \Gamma_{k,2} & \Gamma_{k,1}}.
	\end{align}
\end{widetext}

\bibliography{../../../MK}

\end{document}